\documentclass[aps,pra,reprint,superscriptaddress,amsmath,amssymb]{revtex4-2}

\usepackage{graphicx}
\usepackage{physics}
\usepackage{xr}
\usepackage{verbatim}
\usepackage{color,xcolor}
\graphicspath{{./Figs/}}
\begin{document}
    \title{General Theory for Group Resetting with Application to Avoidance}
    
    \author{Juhee Lee}
    \thanks{These authors contributed equally to this work.}
    \affiliation{Integrated Science Lab, Department of Physics, Ume{\aa} University, SE-90187 Ume{\aa}, Sweden}
    
    \author{Seong-Gyu Yang}
    \thanks{These authors contributed equally to this work.}
    \affiliation{Integrated Science Lab, Department of Physics, Ume{\aa} University, SE-90187 Ume{\aa}, Sweden}
    \affiliation{Department of Physics and Institute of Basic Science, Sungkyunkwan University, Suwon 16419, Republic of Korea}
    
    \author{Hye Jin Park}
    \email[Corresponding author:]{hyejin.park@inha.ac.kr}
    \affiliation{Department of Physics, Inha University, Incheon 22212, Republic of Korea}
    \affiliation{Physics Research Institute, Inha University, Incheon 22212, Republic of Korea}

    \author{Ludvig Lizana}
    \email[Corresponding author:]{ludvig.lizana@umu.se}
    \affiliation{Integrated Science Lab, Department of Physics, Ume{\aa} University, SE-90187 Ume{\aa}, Sweden}
    
    \date{\today}
    
    \begin{abstract}
        We present a general theoretical framework for group resetting dynamics in a potential landscape.
        While traditional resetting models typically focus on a single particle, we consider a group of particles whose collective dynamics govern the resetting.
        We extend existing resetting theories to cover extreme-value group resetting.
        This has applications from bacterial evolution under antibiotic pressure to swarm-search optimization.
        Using renewal theory, we derive a Fokker-Planck equation for the spatial distribution of the group's center of mass, treated as an effective particle.
        This formalism yields analytical expressions for key observables such as the stationary mean position and variance.
        We also study a group avoidance problem, where the particles must avoid an undesirable region.
        Such problems have recently been studied in contexts such as preventing critically high water levels in dams and controlling excessive financial leverage.
        Our framework offers new insight into how resetting can optimize group-level search and avoidance strategies.
    \end{abstract}
    
    \maketitle
    \section{Introduction}
    In traditional stochastic resetting, a particle diffuses through space and relocates to a distant position at a constant rate in order to, for example, find a target~\cite{evans2011diffusion,pal2017first,chechkin2018random,durang2019first}.
    While originally formulated as memoryless, resetting theories now incorporate non-Markovian processes~\cite{harbola2024stochastic,shkilev2017continuous,shkilev2022subdiffusive}, nonequilibrium environments~\cite{maso2019stochastic,goswami2025stochastic}, and bounded or nonrenewal systems~\cite{bodrova2019nonrenewal,mendez2022nonstandard}.
    These developments have proven useful across diverse applications, ranging from transport processes to optimal target search.
    Yet, there exists a class of problems that fall outside the standard single-particle framework: systems with many searchers where the resetting protocol depends on collective behaviors.

    One example is a swarm search~\cite{mesquita2008optimotaxis,kennedy1995particle,wang2018particle}.
    Here, multiple random walkers explore a landscape in parallel, continuously exchanging ``fitness'' values (with respect to some objective function), and, from time to time, relocate to the position of the most fit searcher.
    While this strategy accelerates the search, it raises challenging questions about the optimal number of walkers and the resetting rate.
    Another example is bacterial evolution under antibiotic pressure.
    Under such pressure, the microbial population evolves toward an antibiotic-resistant bacteria through metabolic adaptation~ \cite{davies2010origins,neu1992crisis,munita2016mechanisms,blair2015molecular}.
    This represents an example of group search in biological-trait space.
    Such a process could, in principle, be halted by resetting the population to the least fit species using an artificial selection protocol~\cite{arias2019artificially,sanchez2021directed,thomas2024artificial}.
    Just as in swarm search, finding the optimal parameters (e.g., the best resetting rate) represents a considerable challenge.

    Inspired by these examples, we developed a general framework for group resetting in a drift potential, where the system's current state determines the resetting point.
    It encompasses common resetting rules, such as returning to a fixed position or some spatial distribution~\cite{nagar2023stochastic,basu2019symmetric,sadekar2020zero,karthika2020totally,gupta2014fluctuating,gupta2016resetting,magoni2020ising,durang2014statistical,grange2020non,biroli2023extreme}, but also more complex scenarios where, for example, all particles reset to the location of the most extreme one.
    So far, dynamical resetting rules have mainly been considered in single-particle cases, such as resetting to a fraction of the traveled distance~\cite{tal2022diffusion,di2023time} or to the maximum position previously reached~\cite{majumdar2015random}.

%
    \begin{figure}[t]
    \centering 
    \includegraphics[width=0.9\linewidth]{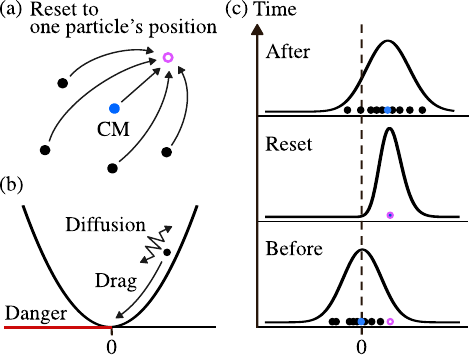} 
    \caption{
        Schematic figure of group resetting.
        (a) Reset dynamics.
        All particles relocate to the position of the selected particle in the group.
        So does the center of mass (CM).
        (b) Avoidance problem.
        The drift potential drags the CM towards the minimum, marking the start of the undesired or dangerous area (red).
        (c) The spatial distributions before and after the group reset (black lines).
        All particles instantaneously relocate to the rightmost particle before resuming diffusion.
    }\label{fig:schematics}
    \end{figure}

    Our framework builds on an effective description in which the system's center of mass (CM) diffuses as a single particle.
    Within this picture, we use renewal theory to derive a master equation for the CM's spatial distribution under group resetting.
    The theory yields several analytical results that agree with many-particle simulations, even under extreme resetting rules.
    We further apply the framework to a generic avoidance problem, calculating the probability that the CM avoids a specified region under counteracting drift.
    This problem arises in diverse contexts~\cite{de2020optimization,de2021optimization}, including preventing dam overflow and controlling excessive financial leverage.
    While these applications rely on single-particle models, our work extends the concept to groups, where avoidance is regulated by resetting to the best-performing member.
    We schematically illustrate the group-resetting process and the avoidance problem in Fig.~\ref{fig:schematics}.

    \section{General framework}
    We consider $n$ overdamped Brownian particles with coordinates $x_i$ ($i=1,\ldots, n$) in a one-dimensional potential $V(x_i)$.
    All particles start at $x_i=x_0$ and diffuse with diffusion constant $D$.
    With resetting rate $r$, the group simultaneously relocates to a position $X(t)$, which in general depends on the particle positions $x_1,\ldots, x_n$.
    To describe the collective dynamics of these particles, we introduce an ``effective particle'' with coordinate $\zeta(t)$ that represents the CM motion.
    It starts at $\zeta_0 = x_0$, evolves with diffusion constant $D/n$ in an effective potential $V_\text{eff}(\zeta)$ (which in general, $V_\text{eff}(\zeta)\neq V(x_i)$), and resets to $\zeta = X(t)$ with rate $r$.
    Following the approach in Ref.~\cite{evans2011diffusion}, we describe the spatial distribution of $\zeta$, $P(\zeta,t|\zeta_0,t_0) \equiv P(\zeta,t)$, by the Fokker-Planck equation
    \begin{equation}\label{eq:FP}
    \begin{split}
    \frac{\partial}{\partial t} P(\zeta,t) &= -\frac{\partial}{\partial \zeta}\left[A(\zeta)P(\zeta,t)\right] +\frac{\partial^2}{\partial \zeta^2}\left[\frac{D}{n}P(\zeta,t)\right]\\
    &-r P(\zeta,t)+ r R(\zeta,t).
    \end{split}
    \end{equation}
    The first two terms represent drift and diffusion, whereas the last two describe removal and reintroduction due to resetting.
    The reintroduction position is drawn from the resetting position distribution $R(\zeta,t)$.

    The main challenge lies in determining $R(\zeta, t)$, since it may depend on the underlying process.
    Specifically, it is determined by the kernel (or conditional probability density function) $K_n(\zeta=X|\zeta'=X';\tau)$, which gives the distribution of the effective particle's resetting position $\zeta=X$ after a time $\tau$ since its last reset to $\zeta'=X'$.
    Following Ref.~\cite{evans2020stochastic}, we employ renewal theory to relate this kernel to $R(\zeta,t)$, obtaining
    \begin{equation}\label{eq:RE_Pr}
    \begin{split}
    R&(\zeta,t) = \Psi(t)K_n(\zeta|\zeta_0;t)\\
    &+\int_0^{t} d\tau \phi(\tau) \int_{-\infty}^\infty d\zeta' K_n(\zeta|\zeta';\tau) R(\zeta',t-\tau).
    \end{split}
    \end{equation}
    Here, $\phi(\tau)$ denotes the waiting-time distribution between consecutive resetting events, $\Psi(t)=1-\int_0^t \phi(\tau)d\tau$ is the survival probability, and the resetting rate is given by $r = \left[\int_0^\infty \tau \phi(\tau)d\tau \right]^{-1}$.
    The first term in Eq.~\eqref{eq:RE_Pr} accounts for contributions from trajectories without resetting up to time $t$, while the second term accounts for trajectories whose most recent reset occurred at time $t-\tau$, irrespective of any earlier resets.
    Together with the initial condition $R(\zeta,0)=\delta(\zeta-\zeta_0)$, Eqs.~\eqref{eq:FP} and~\eqref{eq:RE_Pr}, define the framework for analyzing group resetting dynamics in different scenarios.
    Further progress requires specifying a particular resetting scheme, which leads to different $K_n(\zeta|\zeta';\tau)$.
 
    The simplest example is resetting to the starting or fixed point $\zeta_0$, as in the previous studies~\cite{evans2011diffusion,nagar2023stochastic,biroli2023critical,biroli2023extreme,siboni2021fluctuations,singh2022capture}.
    In this case, $K_n(\zeta|\zeta';\tau) = \delta(\zeta-\zeta_0)$, which yields $R(\zeta,t) = \delta(\zeta-\zeta_0)$.
    Another example is resetting to a randomly chosen particle in the group, for which $K_n(\zeta|\zeta';\tau)=G(\zeta|\zeta';\tau)$, where $G(\zeta|\zeta';\tau)$ denotes the single-particle propagator within time $\tau$.

    A central example in this study is extreme-value group resetting.
    In this scheme, all particles reset to the position of the rightmost one, $X = \max(x_1,\ldots,x_n)$.
    Deriving $K_n(\zeta|\zeta';\tau)$ for this process requires extreme value theory.
    For large $n$, $K_n(\zeta|\zeta';\tau)$ approaches to a Gumbel distribution, $\mathrm{Gumbel}(\zeta;\mu,\beta)$, as predicted by the Fisher-Tippett-Gnedenko theorem~\cite{fisher1928limiting,hansen2020three,cartwright1956statistical}.
    Here, $\mu=\mu(\zeta',\tau)$ and $\beta=\beta(\zeta',\tau)$ are the process-dependent location and scale parameters, respectively.
    A detailed derivation of the approximation $K_n(\zeta|\zeta';\tau) \approx \mathrm{Gumbel}(\zeta; \mu, \beta)$ is provided in the End Matter.

    \section{Results}
%
    \begin{figure}[t]
    \centering 
    \includegraphics[width=\linewidth]{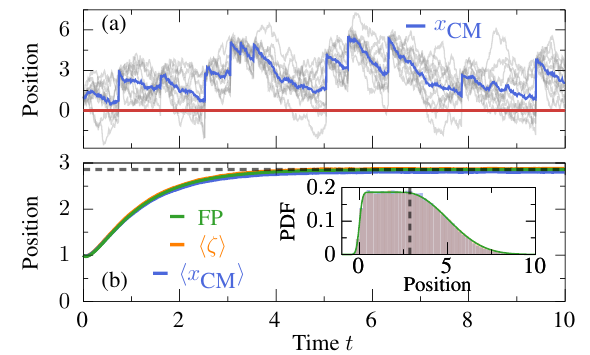}
    \caption{
        Particle trajectories and average positions.
        (a) Simulated particle trajectories with group resetting (grey) and their center of mass (CM, blue).
        The red line marks the boundary of the dangerous region $(x \leq 0)$.
        Parameters: $(n, r, k, D) = (10, 1, 1, 2)$.
        (b) Average positions over time for $n=100$ particles.
        The blue symbols represent the average CM trajectory, and the orange ones show the average from effective particle simulations.
        The green line corresponds to the first moment of $P(\zeta,t)$ from Eq.~\eqref{eq:FP}, and the dashed line indicates the stationary position $\langle\zeta\rangle_s$ [Eq.~\eqref{eq:st_sol}].
        (inset) Position distributions at $t=10$.
        Histograms show simulation results for $\zeta$ and $x_\textrm{CM}$, while the green line is the stationary distribution $P_s(\zeta)$ calculated numerically from Eq.~\eqref{eq:FP}.
        The dashed line shows the first moment of $P_s(\zeta)$.
    }\label{fig:dynamics}
    \end{figure}

    For analytical tractability, we consider the harmonic potential $V(x_i)= k x_i^2 / 2$ [Fig.~\ref{fig:schematics}(b)].
    This corresponds to an Ornstein-Uhlenbeck process~\cite{uhlenbeck1930theory, gardiner2009stochastic}, which allows us to determine the parameters $\mu$ and $\beta$ analytically from the single-particle propagator.
    The location and scale parameters are $\mu(\zeta',\tau) = \overline x(\zeta',\tau)+\sigma(\zeta',\tau) b_n$ and $\beta(\zeta',\tau) = \sigma(\zeta',\tau) a_n$, where $\overline{x}$ and $\sigma^2$ indicate the mean and variance of the Ornstein-Uhlenbeck process, $\overline x(\zeta',\tau)=\zeta'e^{-k\tau}$ and $\sigma^2(\zeta',\tau)=\frac{D}{k}[1 - \exp(-2k\tau)]$.
    Here, the coefficients $b_n$ and $a_n$ are determined by the inverse cumulative distribution $C^{-1}(\cdot)$ of a unit Gaussian distribution, where $b_n=C^{-1}(1-1/n)$ and $a_n=1/b_n$.

    As a basic of avoidance, we first examine the first moment as it represents the mean displacement from the danger boundary $(\zeta=0)$.
    To this end, we compare the dynamics of the CM with our effective particle description under resetting.
    Figure~\ref{fig:dynamics}(a) shows the trajectories of $n=10$ diffusing particles (grey), which reset at rate $r=1$ to the position of the rightmost particle.
    The CM trajectory $x_{\text{CM}} (= \frac{1}{n}\sum_{i=1}^{n}x_i)$ is colored in blue, where the sudden jumps indicate resetting events.

    Figure~\ref{fig:dynamics}(b) shows the ensemble-averaged CM motion, $\langle x_{\text{CM}}(t)\rangle$ ($10^5$ samples), together with the average trajectory of the effective particle, $\langle \zeta \rangle$ (orange), obtained from stochastic simulations~\cite{kloeden1992stochastic} (also $10^5$ samples, see End Matter).
    Both averages are nearly identical and saturate in the long time limit, $\langle x_{\text{CM}}(t\to \infty)\rangle\approx\langle \zeta (t\to\infty) \rangle$.
    These results demonstrate that the effective-particle description provides an accurate representation of the group's CM.

    In addition to the simulations in Fig.~\ref{fig:dynamics}(b), we show the theoretical prediction of the first moment $\langle \zeta(t)\rangle$ (green), obtained by solving the dynamics of $\zeta(t)$.
    The dynamics of $\langle \zeta(t) \rangle$ is obtained from Eqs.~\eqref{eq:FP} and \eqref{eq:RE_Pr} by multiplying $\zeta$ by $P(\zeta,t)$ and $R(\zeta, t)$ and integrating over $\zeta$, which yields
    \begin{equation}\label{eq:mean_zeta}
        \frac{d}{dt}\langle\zeta(t)\rangle = -k\langle\zeta(t)\rangle + r\left(\langle\zeta_r(t)\rangle - \langle \zeta(t) \rangle \right),
    \end{equation}
    where
    \begin{equation}\label{eq:mean_zetar}
    \begin{split}
        \langle \zeta_r&(t)  \rangle = \int_{-\infty}^\infty \zeta R(\zeta,t)d\zeta \\
        &=\Psi(t)\left[\zeta_0e^{-kt}+(b_n+\gamma a_n) \sqrt{\frac{D}{k} \left( 1 - e^{-2kt} \right) }\right]\\
        &\quad +\int_0^{t} d\tau \phi(\tau) \left[ \langle \zeta_r(t-\tau) \rangle e^{-k\tau}   \vphantom{\sqrt{\frac{D}{k}} }\right.\\
        &\quad \left.+(b_n+\gamma a_n) \sqrt{ \frac{D}{k} \left( 1 - e^{-2k\tau} \right) } \right].
    \end{split}
    \end{equation}
    Here, $\gamma$ is the Euler-Mascheroni constant.
    In Eq.~\eqref{eq:mean_zeta}, the first term describes the drift toward $\zeta=0$, while the last term represents the resetting contribution, which counteracts this drift by displacing the effective particle away from $\zeta=0$.

    Using $\phi(t) = r e^{-r t}$, we obtain the solution to Eq.~\eqref{eq:mean_zeta} [green in Fig.~\ref{fig:dynamics}(b)], which is in excellent agreement with the simulations (blue and orange).
    To further validate our theory, we compared the stationary solution of Eq.~\eqref{eq:FP} (green), obtained by numerical integration, with histograms of the stationary positions $\zeta$ (orange) and $x_{\text{CM}}$ (blue) from simulations [Fig.~\ref{fig:dynamics}(b)~(inset)].
    Again, we find good agreement.

    Next, the effective particle description allows us to calculate several analytical results, including the stationary mean position $\langle\zeta\rangle_s= \lim_{t\to \infty}\langle\zeta(t)\rangle$.
    This is a simple proxy for avoidance given by
    \begin{equation}\label{eq:st_sol}
        \langle\zeta\rangle_s = \left(b_n + \gamma a_n \right) \sqrt{\frac{D}{k}} \frac{r^2}{2k(r+k)} B\left(\frac{1}{2}, \frac{r}{2k}\right),
    \end{equation}
    where $B(u,v) = \Gamma(u)\Gamma(v)/\Gamma(u+v)$ denotes the beta function; $\Gamma(\cdot)$ is the gamma function.
    Below, we examine how Eq.~\eqref{eq:st_sol} depends on key parameters, starting with the group size $n$.

    \begin{figure}[t]
    \centering 
    \includegraphics[width=\linewidth]{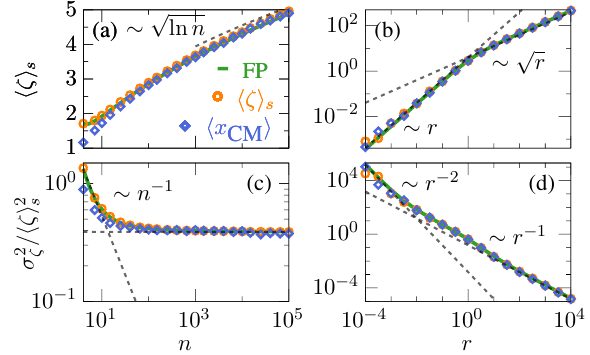} 
    \caption{
        [(a) and (b)] The stationary mean position $\langle \zeta \rangle_s$ and [(c) and (d)] the squared coefficient of variation (SCV) $\sigma^2_\zeta/\langle\zeta\rangle_s^2$ with respect to [(a) and (c)] $n$ for $(r,k,D) = (1,1,2)$, and [(b) and (d)] $r$ for $(n,k,D)=(100,1,2)$.
        $\langle \zeta \rangle_s$ increases with $n$ and $r$, while the SCV decreases for both parameters.
        However, the SCV saturates at large $n$.
        The blue and orange symbols indicate simulation results for groups of particles and the effective center of mass, respectively.
        The green lines represent theoretical results, and the dashed lines depict scaling trends.
    }\label{fig:stationary}
    \end{figure}

    The size is essential in group resetting, as it determines the group's extreme value statistics: the bigger the groups, the more extreme the outcomes.
    As $n$ grows, this leads to a growing $\langle \zeta \rangle_s$.
    By expanding Eq.~\eqref{eq:st_sol} and using $b_n \propto \sqrt{2\ln n}$ for large $n$~\cite{fisher1928limiting,hansen2020three,cartwright1956statistical}, we find $\langle \zeta \rangle_s \propto \sqrt{\ln n}$ [Fig.~\ref{fig:stationary}(a)]. 
    This growth is slow.
    The opposite small-$n$ limit is relatively trivial.
    When $n=1$, the particle has nowhere to jump since there are no interactions with other particles.
    Therefore, $\langle\zeta\rangle_s = 0$.

    Next, we study the impact of resetting rate $r$.
    Without resetting ($r=0$), we have $\langle \zeta \rangle_s = 0$, because the effective particle simply follows the Ornstein-Uhlenbeck process.
    As $r$ grows, however, $\langle\zeta\rangle_s$ increases [Fig.~\ref{fig:stationary}(b)].
    We note that the system exhibits two scaling regimes for small and large $r$.
    We analyze the small-$r$ regime ($r \ll k$) by expanding Eq.~\eqref{eq:st_sol} and using that $B(1/2, r/2k) \approx 2k/r - 2r/k$.
    This leads to the linear scaling:
    \begin{equation}\label{eq:small_r}
        \langle\zeta\rangle_s \approx (b_n+\gamma a_n) \frac{r}{k}\sqrt{\frac{D}{k}}.
    \end{equation}
    For large $r\, (r \gg k)$, we find $\langle \zeta \rangle_s \propto \sqrt{r}$ by expanding Eq.~\eqref{eq:st_sol} and using $B(1/2, r/2k) \sim \sqrt{2\pi k/r}$,
    \begin{equation}\label{eq:large_r}
        \langle\zeta\rangle_s \approx ( b_n + \gamma a_n ) \sqrt{\frac{\pi D r}{2 k^2}}.
    \end{equation}
    This limiting behavior has an intuitive explanation.
    During the time interval between two resets, $\tau\approx 1/r$, diffusive spread grows as $\sqrt{D/r}$ and the particle drifts toward the origin by a distance of $k\zeta /r$.
    Under stationary conditions, these two mechanisms balance each other, $k\langle\zeta\rangle_s /r \sim \sqrt{D/r}$, which yields $\langle \zeta \rangle_s \propto \sqrt{Dr}/k$.

    As mentioned before, the average stationary position is a simple measure for avoidance, where we show that it increases with both $n$ and $r$.
    At first glance, one might interpret that the group successfully avoids the danger in the stationary state if $\langle\zeta\rangle_s>0$.
    However, this is not guaranteed, because the group may fail to avoid the danger if the variance $\sigma_\zeta^2\,\left(= \langle \zeta^2 \rangle_s - \langle \zeta \rangle_s^2 \right)$ grows faster than $\langle\zeta\rangle_s$ with $n$ and $r$.
    To capture this behavior, we calculate the squared coefficient of variation (SCV) of the displacement $\sigma_\zeta^2 / \langle \zeta \rangle^2_s$ as a better dimensionless measure of avoidance.

    To evaluate the SCV, we must first calculate the second moment in steady-state $\langle\zeta^2\rangle_s= \lim_{t\to \infty} \langle \zeta^2(t)\rangle$.
    In the same manner as for $\langle \zeta \rangle_s$, we find
    \begin{equation}\label{eq:st2_sol}
    \begin{split}
        \langle \zeta^2 &\rangle_s = \frac{2D}{n(r+2k)} + \left[ (b_n + \gamma a_n)^2 + \frac{\pi^2 a_n^2}{6} \right] \frac{ Dr}{k(r+2k)}\\
        &+\frac{(b_n + \gamma a_n)^2 D r^3}{4k^3(r+2k)} B\left(\frac{1}{2}, \frac{r}{2k}\right)B\left(\frac{1}{2}, \frac{r+k}{2k}\right).
    \end{split}
    \raisetag{18pt}
    \end{equation}
    Using this formula, we illustrate how the SCV changes for different $n$ and $r$ in Fig.~\ref{fig:stationary}(c) and (d).

    First, we find that the SCV decreases as $\sim 1/n$ for small $n$, and then saturates for large $n$ [Fig.~\ref{fig:stationary}(c)].
    This indicates that the stationary distribution $P_s(\zeta)$ broadens proportionally to the shift of the average position for large $n$.
    Second, with respect to $r$, the SCV decreases monotonically, but with two distinct regimes.
    It scales as $\sim 1/r^2$ for small $r\, (\ll k)$, while for large $r\, (\gg k)$, it decays as $1/r$ [Fig.~\ref{fig:stationary}(d)].
    This monotonic decrease in SCV with increasing $r$ arises from the fact that the fast resetting shortens the diffusion time, thereby limiting the broadening of the distribution.
    These results imply that the larger the $n$ and the larger the $r$, the more likely the group is to avoid the danger.
    We also find that $\langle\zeta\rangle_s$ and the SCV are more sensitive to $r$ than $n$.

    \begin{figure}[t]
    \centering 
    \includegraphics[width=\linewidth]{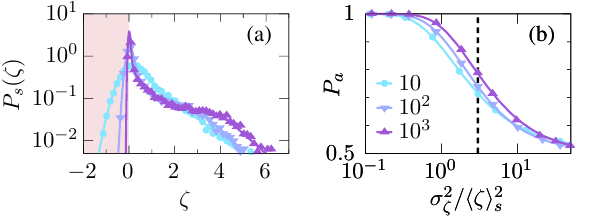} 
    \caption{
        (a) Stationary distributions $P_s(\zeta)$ for different parameter values of $n$ and $r$.
        These distributions share the same squared coefficient of variation (SCV) value, $\sigma_\zeta^2 / \langle \zeta \rangle^2_s = 3$, but differ in their detailed shapes.
        Red area represents the dangerous region.
        (b) Avoidance probability $P_a$ as a function of SCV for different group size $n$.
        Symbols represent simulation results, and lines represent the numerical solution of Eq.~\eqref{eq:FP}.
        The dashed line indicates $\sigma_\zeta^2 / \langle \zeta \rangle^2_s = 3$.
        The parameters are $(k,D)=(1,2)$.
    }\label{fig:avoidance}
    \end{figure}

    Interestingly, although the SCV is an improved measure of avoidance relative to the average displacement, two groups with the same SCV but with different parameter combinations may exhibit different avoidance behavior.
    This is because avoidance ultimately is a function of the detailed shape of the distribution $P_s(\zeta)$, not just its mean and variance.
    Figure~\ref{fig:avoidance}(a) shows three stationary distributions $P_s(\zeta)$ for different parameters $(n,r)$ that yield the same SCV value, $\sigma_\zeta^2 / \langle \zeta \rangle_s^2 = 3$.
    The lines represent results from numerical solutions of Eq.~\eqref{eq:FP}, and the symbols are computed from numerical integration of the stochastic differential equation for $\zeta$.
    Despite having the same SCV, the distributions exhibit different shapes.

    To get the proper avoidance probability, we calculate $P_a=\int_0^\infty d\zeta P_s(\zeta)$, which represents the probability of finding the effective particle away from the dangerous region $\zeta\leq 0$.
    In Fig.~\ref{fig:avoidance}(b), we plot $P_a$ versus SCV for different $n$.
    Successful avoidance $(P_a = 1)$ is achieved when the distribution is entirely shifted to the right of the origin, ensuring that no portion of it remains in $\zeta\leq0$.
    This implies that the mean is shifted farther than the width of the distribution, such that successful avoidance is associated with a small SCV.

    \section{Summary and discussion}
    In summary, we present a general theoretical framework for group resetting that builds on renewal theory and can handle extreme value scenarios.
    We derive the renewal equation for the resetting position distribution $R$ and couple it to a Fokker-Planck equation that governs the group's center of mass dynamics.
    We also derive several analytical results for the stationary mean position and show how key resetting and group parameters control the system's ability to avoid an undesirable area.

    Reset protocols in stochastic resetting are typically categorized into two classes.
    One is global resetting, where all particles reset simultaneously~\cite{evans2020stochastic,nagar2023stochastic,evans2022exactly,mercado2018lotka,miron2021diffusion,biroli2023extreme,biroli2023critical,singh2022capture,singh2023bernoulli}, while the other is local resetting, where particles reset independently~\cite{nagar2023stochastic,da2018interplay,grange2020non,vilk2022fluctuations,krapivsky2022competition,siboni2021fluctuations,biroli2023critical,sasorov2023probabilities}.
    The extreme-value group resetting in our study falls within the global resetting class.

    Beyond this classification, however, a more crucial distinction concerns how the resetting position is determined.
    Most previous studies have considered particles, either globally or individually, resetting to a predefined state, such as the origin~\cite{evans2020stochastic,nagar2023stochastic,evans2022exactly,mercado2018lotka,da2018interplay,miron2021diffusion,biroli2023extreme,sasorov2023probabilities} or their initial state~\cite{evans2020stochastic,nagar2023stochastic,grange2020non,biroli2023critical,singh2022capture,singh2023bernoulli}.
    In contrast, in our example the resetting position emerges dynamically from the collective group evolution between consecutive resetting events.
    As a result, the resetting position is not prescribed a priori, but is instead defined collectively by the system dynamics.

    At a conceptual level, our model bears a superficial resemblance to branching Brownian motion~\cite{siboni2021fluctuations} in that collective behavior emerges without direct particle interactions.
    However, the underlying mechanisms are different.
    In branching Brownian motion, collective effects arise from reproduction or birth event, with new particles appearing at positions determined solely by individual trajectories.
    As a result, collective behavior emerges only at the level of ensemble statistics.
    In the extreme-value resetting case, the resetting position is explicitly constructed from the group configuration, through the position of the farthest particle, thereby inducing collective effects even though the underlying Brownian dynamics of individual particles remain independent.

   The framework is broadly applicable to a wide range of group resetting problems.
    By appropriately defining the resetting position distribution $R$ and the kernel $K$, it connects studies on fixed resetting distributions~\cite{evans2011diffusion,olsen2023steady,mori2023entropy,mendez2024first,dahlenburg2021stochastic}, position-dependent resetting~\cite{tal2022diffusion,di2023time,dahlenburg2021stochastic}, simultaneous group resetting to the origin~\cite{biroli2023extreme}, and time-dependent resetting based on particle trajectory history~\cite{boyer2017long,majumdar2015random}.

    Building on this theoretical framework, our approach could be extended to practical applications such as artificial selection~\cite{Lee_2025}.
    For example, to design optimal selection protocols, reduce computational costs, and estimate key parameters such as bottleneck size, fitness, and selection frequencies.
    Resetting tied to group thresholds may also offer new insights into allele frequencies and mechanisms that help populations avoid extinction.

%
%
    \section*{Acknowledgments}
    J.L., S.-G.Y., and L.L. acknowledge financial support from the Swedish Research Council (Grant No.~2021-04080).
    J.L and S.-G.Y are supported by postdoctoral fellowships from the Kempestiftelserna (Grant No.~JCK22-0026.3) and the Carl Tryggers Stiftelse f{\"o}r Vetenskaplig Forskning (Grant No.~CTS 22:2243), respectively.
    H.J.P is supported by the National Research Foundation of Korea grant funded by the Korean government (MSIT) (Grant No.~RS-2023-00214071, RS-2023-NR075951, and RS-2024-00460958).
    This work is also supported by the Swedish Foundation for International Cooperation in Research and Higher Education (STINT) (Grant No.~MG2022-9405) and Basic Science Research Program through the National Research Foundation of Korea (NRF) funded by the Ministry of Education (Grant No.~RS-2025-02312897).

    \section*{Data Availability}
    The data that support the findings of this article are openly available~\cite{our_git}.

    \appendix

%
%
    \section{Derivation of kernel}\label{appendix:a}
    \setcounter{equation}{0}
    \setcounter{secnumdepth}{2}
    \renewcommand{\theequation}{A\arabic{equation}}
    Let us assume that a group of particles has reset to $\zeta'$ at time $t'$, and the next reset occurs at time $t\,(=t'+\tau)$ to $\zeta$.
    To obtain the kernel $K_n(\zeta|\zeta';\tau)$, we start from the cumulative distribution function $Q_n(\zeta|\zeta';\tau) = \textrm{Pr}\left(x_1(t),\ldots, x_n(t) \leq \zeta|\zeta';\tau\right)$.
    Assuming that the particles diffuse independently between successive resets, the cumulative distribution function factorizes as $Q_n(\zeta|\zeta';\tau) = \left[\textrm{Pr}\left( x(t) \leq \zeta|\zeta'; \tau \right) \right]^n,$ where the single-particle cumulative probability is given by $\textrm{Pr}(x \leq \zeta|\zeta';\tau) =\int_{-\infty}^\zeta dx\, G(x|\zeta';\tau) = C(\zeta|\zeta',\tau)$, with the single-particle propagator $G(x|\zeta';\tau)$ from $\zeta'$ to $x$ over a time interval $\tau$, assuming time-homogeneous diffusion.
    Here, $C(\cdot)$ is the cumulative distribution of the unit Gaussian distribution.
    Differentiating $Q_n$ with respect to $\zeta$, we obtain
    \begin{equation} \label{eq:resetting_distr}
    K_n(\zeta|\zeta'; \tau) = n G(\zeta|\zeta';\tau)\left[C(\zeta|\zeta';\tau)\right]^{n-1}.
    \end{equation}

    As $n$ increases, $K_n(\zeta|\zeta';\tau)$ approaches a Gumbel distribution (or type I generalized extreme value distribution) $\textrm{Gumbel}(\zeta; \mu, \beta)= \frac{1}{\beta} \exp\left(-\frac{\zeta-\mu}\beta - \exp\left(-\frac{\zeta-\mu}\beta\right)\right)$ as stated in the Fisher-Tippett-Gnedenko theorem, since the single-particle propagator has an exponential tail~\cite{fisher1928limiting,hansen2020three,cartwright1956statistical}.
    Parameters $\mu$ and $\beta$ represent the location and scale of the Gumbel distribution, respectively.
    Such emergence of the Gumbel distribution is a hallmark of extreme value theory, reflecting the fact that the group resetting process selects the maximal displacement among diffusing particles.

    We now demonstrate this relation explicitly, without invoking the Fisher-Tippett-Gnedenko theorem.
    Starting from the single-particle Gaussian propagator $G(x)$, we scale the variable as $y=(x-\overline{x})/\sigma$, where $\overline{x}=\zeta'e^{-k\tau}$ and $\sigma=\sqrt{\frac{D}{k}\left(1-e^{-2k\tau}\right)}$, which gives $G(y) = \frac{1}{\sqrt{2\pi}} e^{-y^2/2}$.
    In this setting, the asymptotic order $b_n$ of the maximum follows from $\textrm{Pr}(y > b_n) = 1-C(b_n) \sim 1/n$.
    This implies that $b_n \approx C^{-1}(1-1/n)$.
    For simplicity, we omit the dependence on $\zeta'$ and $\tau$ in what follows.

    To evaluate $b_n$, we use the Mills ratio $m(b_n)$ for the unit Gaussian,
    \begin{equation} \label{eq:mills_ratio}
        m(b_n) = \frac{1-C(b_n)}{G(b_n)} 
           = \frac{1}{b_n + \frac{1}{b_n + \frac{2}{b_n + \frac{3}{b_n + \cdots}}}},
    \end{equation}
    which reduces to $m(b_n) \approx 1/b_n \left(1 +O(b_n^{-2}) \right)$ for large $b_n$, giving $C(b_n) \approx 1-G(b_n)/b_n$.
    Hence, $b_n \approx \sqrt{\ln(n^2/2\pi) - \ln[\ln(n^2/2\pi)]}$.

    As the group size $n$ increases, the order $b_n$ of maximum also increases.
    Simultaneously, the tail of the distribution $[G(y)]^n$ for $n$ independent random variables decays more steeply.
    With $b_n$ identified, we rescale the variables $y_i$ as $z_i = (y_i - b_n)/a_n$ with $a_n=1/b_n$ to resolve the tail behavior.
    The distribution satisfies $\textrm{Pr}(z_1,z_2,\ldots,z_n \leq Z) = \left[\textrm{Pr}\left(z \leq Z\right) \right]^n=\left[ C(b_n + a_n Z) \right]^n$.
    Using the Mills ratio, the cumulative distribution can be approximated as $C(b_n+a_n Z )\approx 1-G(b_n+a_n Z)/(b_n + a_n Z)$.
    Moreover, the asymptotic relation $G(b_n+a_n Z) \approx e^{-Z}G(b_n) \left( 1 + O(b_n^{-2})\right)$ yields $C(b_n+a_n Z) \approx 1-e^{-Z}/n$, since $G(b_n)/b_n \approx 1/n$.
    Thus, in the large-$n$ limit, $ \left[\textrm{Pr}\left(z \leq Z\right) \right]^n \approx \left( 1 - \frac{e^{-Z}}{n} \right)^n \rightarrow e^{-e^{-Z}}$, which is the cumulative distribution of the Gumbel distribution.
    
    In terms of $x$, the cumulative distribution becomes
    \begin{equation} \label{eq:Cumul_Gumbel}
        Q_n(\zeta) = \left[ \textrm{Pr}(x\leq \zeta)\right]^n \approx \exp\left( -\exp\left(- \frac{\zeta-\mu}{\beta} \right) \right),
    \end{equation}
    where $\mu=\overline{x}+\sigma b_n$ and $\beta=\sigma a_n$.
    Differentiating $Q_n(\zeta)$ with respect to $\zeta$, we obtain the kernel for large $n$ as
    \begin{equation} \label{eq:Kernel_Gumbel}
        K_n(\zeta) \approx \frac{1}{\beta}\exp\left(-\frac{\zeta-\mu}{\beta}-\exp\left(-\frac{\zeta-\mu}{\beta}\right) \right).
    \end{equation}
    Thus, for large $n$, the kernel is well approximated by the Gumbel distribution.

    \section{Stochastic simulations}\label{appendix:b}
    \setcounter{equation}{0}
    \renewcommand{\theequation}{B\arabic{equation}}
    We generate trajectories using the Euler-Maruyama method~\cite{kloeden1992stochastic,gardiner2009stochastic}.
    At each time step of size $dt$, the positions $(x_1, x_2, \ldots, x_n)$ of $n$ particles are reset with probability $r dt$.
    When a reset occurs, all particles are relocated simultaneously to $X=\max(x_1,x_2,\ldots,x_n)$.
    Otherwise, the particles evolve independently according to
    \begin{equation}\label{group_sde}
        x_i(t+dt) = x_i(t)-k x_idt+\sqrt{2D}\ dW_i(t),
    \end{equation}
    where $W_i(t)$ denotes a Wiener process.

    When simulating the effective single particle dynamics with coordinate $\zeta$, we keep track of the elapsed time $\tau$ since the most recent reset.
    At each time step $dt$, a reset occurs with probability $r dt$.
    In such case, $\zeta$ is relocated to $\zeta_r$, randomly drawn from $K_n(\zeta|\zeta';\tau)$.
    For practical implementation, we approximate $K_n(\zeta|\zeta'; \tau)$ by a Gumbel distribution.
    Between resetting events, $\zeta$ evolves according to
    \begin{equation}\label{cm_sde}
        \zeta(t+dt) = \zeta(t)-k \zeta dt+\sqrt{2D/n}\ dW_\zeta(t),
    \end{equation}
    where $W_\zeta(t)$ is also a Wiener process.

    \section{Analytic expression of squared coefficient of variation}\label{appendix:c}
    \setcounter{equation}{0}
    \renewcommand{\theequation}{C\arabic{equation}}

    An explicit expression for the squared coefficient of variation can be obtained as
    \begin{equation}
    \begin{split}
    \frac{\sigma^2_\zeta}{\langle\zeta\rangle_s^2}&=\frac{8k^{3}(r+k)^{2}}{nr^{4}(r+2k)}\left(b_{n}+\gamma a_{n}\right)^{-2} B\left(\frac{1}{2},\frac{r}{2k}\right)^{-2}
        \\
        &+\left[1+\frac{\pi^{2}a_{n}^{2}}{6\left(b_{n}+\gamma a_{n}\right)^{2}}\right]\frac{4k^{2}(r+k)^{2}}{r^{3}(r+2k)B\left(\frac{1}{2},\frac{r}{2k}\right)^{2}}
        \\
        &+\frac{(r+k)^{2}}{r(r+2k)}\frac{B\left(\frac{1}{2},\frac{r+k}{2k}\right)}{B\left(\frac{1}{2},\frac{r}{2k}\right)}-1.
    \end{split}
    \end{equation}
    \bibliographystyle{apsrev4-2}
    \bibliography{ref}
\end{document}